\documentclass[aps,reprint,superscriptaddress]{revtex4-1}

\usepackage{graphicx} 
\usepackage{amsmath,bm}	
\begin{document}
\title{Finite-size scaling with respect to interaction and disorder strength at the many-body localization transition} 

\author{Kazue Kudo}
\email[]{kudo@is.ocha.ac.jp}
\affiliation{Department of Computer Science, Ochanomizu University,  
Tokyo 112-8610, Japan}

\author{Tetsuo Deguchi}
\affiliation{Department of Physics, Ochanomizu University,  
Tokyo 112-8610, Japan}

\date{\today}

\begin{abstract}
We present a finite-size scaling for both interaction and disorder strengths in the critical regime of the many-body localization (MBL) transition for a spin-1/2 XXZ spin chain with a random field by studying level statistics.  
We show how the dynamical transition from the thermal to MBL phase depends on interaction together with disorder by evaluating the ratio of adjacent level spacings, 
and thus, extend previous studies in which interaction coupling is fixed.  
We introduce an extra critical exponent in order to describe the nontrivial interaction dependence of the MBL transition.  
It is characterized by the ratio of the disorder strength to the power of the interaction coupling with respect to the extra critical exponent and not by the simple ratio between them.
\end{abstract}

\maketitle

\section{Introduction}

Recently, many-body localization (MBL) has attracted much interest in many-body quantum physics and statistical mechanics~\cite{basko2006,znidaric2008,monthus2010,canovi2011,canovi2012,bardarson2012,serbyn2013a,serbyn2013b,huse2014,pal2010,serbyn2016,luitz2015,oganesyan2007,bertrand2016,khemani2017a,khemani2017b,enss2017}. 
In the MBL phase, the dynamics is non-ergodic and the system does not thermalize, which implies that the eigenstate thermalization hypothesis (ETH)~\cite{deutsch1991,srednicki1994,rigol2008} is broken.
The MBL phase has similarities with integrable systems in the sense that eigenstates can be characterized by a large number of local integrals of motion~\cite{serbyn2013b,chandran2015,ros2015,luitz2015}.
The transition between the ETH and MBL phases is characterized by several quantities such as level statistics, entanglement entropy, the Kullback-Leibler divergence, and bipartite fluctuations of subsystem magnetization (in a spin chain).
For instance, intensive numerical simulations on those quantities~\cite{luitz2015} 
illustrate that the MBL transition points characterized by them coincide with each other.

Numerical analysis of spectral properties may provide fundamental information on  many-body quantum systems. 
We often investigate the level statistics of a quantum system in order to determine whether it is integrable.  
If a given Hamiltonian is integrable by the Bethe ansatz, level statistics obeys the Poisson distribution, while if it is nonintegrable, the Wigner-Dyson (WD) statistics is yielded. 
This has been confirmed in correlated quantum spin systems~\cite{montambaux1993,hsu1993,poilblanc1993,vanede1994,meyer1996,meyer1997_1,meyer1997_2,dauriac1998,dauriac2003,rabson2004,kudo2003,kudo2005,santos2010,santos2012,gubin2012,vahedi2016} 
as well as disordered systems~\cite{berkovits1996,georgeot1998,avishai2002,santos2004,kudo2004,brown2008,santos2018}.
The WD and Poisson statistics also characterize the metal and insulator (localized) phases, respectively, in the Anderson model of disordered systems~\cite{shklovskii1993}.
MBL generalizes Anderson localization (AL).
In disordered quantum many-body systems, level statistics changes from the WD to Poisson statistics if we increase disorder. 
The transition was found in spin systems more than a decade ago~\cite{santos2004,kudo2004}, and they are regarded as some of the earliest examples of MBL transitions. 

MBL transitions are usually studied in terms of disorder strength under a fixed interaction. 
It has not been discussed intensely, yet, how it depends on interaction coupling.   
In this paper, we will demonstrate that the MBL transition indeed depends on interaction in a nontrivial manner.  
We present a finite-size scaling of level statistics for both interaction and disorder in the critical regime.  
The nonlinear interaction dependence is characterized by the ratio of the disorder strength to some power of the interaction coupling, 
and we introduce a critical exponent for the power.  
The finite-size scaling of level statistics with the extra critical exponent should be useful in further studies of finite-size scaling in the MBL transition such as that of entanglement entropy.  
In fact, level statistics requires much less computational costs than other observables such as entanglement entropy, since it only needs eigenenergies, and it works sufficiently well for characterizing MBL phases. 

In this paper, we investigate level statistics for a spin-1/2 XXZ spin chain with a random field.
We introduce the model and methods in Sec.~\ref{sec:model}.
In Sec.~\ref{sec:dependence}, how the level statistics depends on disorder and interaction is demonstrated in the gapful regime as well as the gapless regime.
Next, we focus on the MBL transition in the gapless regime and perform a finite-size scaling analysis in Sec.~\ref{sec:scaling}.
Discussion and conclusions are given in Sec.~\ref{sec:discussion}.

\section{Model and methods}
\label{sec:model}

Let us consider a spin-1/2 XXZ spin chain with a random magnetic field.
The Hamiltonian on $L$ sites is given by
\begin{equation}
 H = \sum_{j=1}^L\left(
S_j^x S_{j+1}^x + S_j^y S_{j+1}^y + \Delta S_j^z S_{j+1}^z 
\right) + \sum_{j=1}^L h_j S_j^z,
\label{eq:H}
\end{equation}
where $S^\alpha=\frac12\sigma^\alpha$ and $\sigma^\alpha$ with $\alpha=x,y,z$ are the Pauli matrices;
$\Delta$ is the anisotropy parameter; 
$h_j$ is a random magnetic field at site $j$ with a uniform distribution $[-h:h]$; 
the periodic boundary conditions are imposed.
The XXZ spin chain can be mapped into interacting spinless fermions in a one-dimensional lattice.
In the system of spinless fermions, anisotropy $\Delta$ corresponds to the interaction between such fermions that are located on neighboring sites.
We hereafter refer to $\Delta$ and $h$ as interaction and disorder, respectively. 

If the smallest energy necessary for the ground state to be excited remains nonzero and constant as the system size $L$ goes to infinity,  we say that the spectrum has a gap or it is gapful,  
while if it is proportional to $1/L$ or smaller than that as $L$ goes to infinity, we say that the spectrum has no gap or it is gapless. When all the random fields $h_j$ vanish in  the Hamiltonian (\ref{eq:H}), its spectrum  is gapful for $|\Delta| > 1$ and gapless for  $|\Delta| \le 1$.    

We employ the ratio of adjacent level spacings,
which was recently introduced in order to study level statistics~\cite{oganesyan2007,luitz2015,bertrand2016,khemani2017b,khemani2017a}.
The ratio is defined by
\begin{align}
 r_i &=\frac{\min (\delta_i,\delta_{i+1})}{\max (\delta_i,\delta_{i+1})},
\label{eq:r}
\\
\delta_i &= E_i-E_{i-1},
\end{align}
where $E_i$ is the $i$-th eigenvalue (in ascending order) of a given energy spectrum, and hence, 
$\delta_i$ is the level spacing between the $i$-th and ($i-1$)-th eigenvalues.
The average values of the ratio for the Poisson and WD distributions are given by 
$\langle r\rangle_{\rm p}=2\ln 2-1 \approx 0.386$ and 
$\langle r\rangle_{\rm w} \approx 0.530$, respectively~\cite{oganesyan2007}.
In many studies of level statistics, it is common to use unfolded eigenvalues instead of raw eigenvalues.
The unfolding procedure is given by a method for rescaling the energy spectrum so that the local level density becomes unity.
We make use of unfolded spectra in order to calculate the average ratio $\langle r\rangle$, which is actually independent of the unfolding procedure.
Here, we follow the unfolding procedure (specifically, the average cubic global version) proposed in Ref.~\onlinecite{bertrand2016} .

 \begin{table}
 \caption{\label{tab:number} 
The number of level spacings taken in each spectrum and the number of disorder realizations for each system size $L$.}
 \begin{ruledtabular}
 \begin{tabular}{crr}
 $L$ & No. level spacings & No. realizations\\
\hline
  12 & 100 & 10000\\
  13 & 100 & 10000\\
  14 & 500 & 2000\\
  15 & 500 & 2000\\
  16 & 1000 & 1000\\
 \end{tabular}
 \end{ruledtabular}
 \end{table}

The system size considered below is given by $L=12$, $13$, $14$, $15$, and $16$.
Since the eigenvalues with different total $S^z$ are uncorrelated,
we consider only the largest subspaces: $S^z=0$ for even numbers of $L$ and $S^z=1$ for odd ones.
Eigenvalues were computed with the {\sc lapack} library.
Spectral properties depend on the energy range~\cite{luitz2015,bertrand2016}.
It is safe to take only the central part in each spectrum.
The energy range of each spectrum used below is about 6\%--15\% of the full spectrum.
The number of disorder realizations and the number of level spacings taken in each realization are listed in Table~\ref{tab:number}.

\section{Dependence on disorder and interaction}
\label{sec:dependence}

\begin{figure}
\includegraphics[width=8cm,clip]{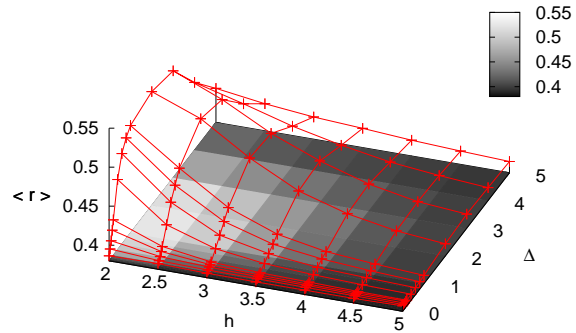}%
\caption{\label{fig:map} (Color online)  
The average ratio $\langle r \rangle$ of adjacent level spacings for $L=16$ shows dependence on disorder $h$ and interaction $\Delta$.
The Poisson and Wigner-Dyson distributions correspond to $\langle r\rangle_{\rm p} \approx 0.386$ and $\langle r\rangle_{\rm w} \approx 0.530$, respectively.
Error bars, which are smaller than $10^{-3}$, are omitted.
For $\Delta\neq 0$, average ratio $\langle r\rangle$ decreases as $h$ increases, which indicates the MBL transition.
At $\Delta=0$, $\langle r\rangle\approx\langle r\rangle_{\rm p}$ for an arbitrary $h$, which corresponds to Anderson localization.
}
\end{figure}

Figure~\ref{fig:map} illustrates average ratio $\langle r\rangle$ for $L=16$ as a function of disorder $h$ and interaction $\Delta$.
Average ratio $\langle r\rangle$ decreases with increasing $h$ (for $\Delta\neq 0$), which indicates the MBL transition.
The ratio also depends on interaction: 
it increases with interaction for $\Delta\lesssim 2$ and decreases for $\Delta\gtrsim 2$.
This is consistent with the dynamical phase diagram for one-dimensional spinless fermions with random fields~\cite{lev2015, bera2015}, where the system can be mapped to the spin-1/2 XXZ spin chain with random fields.
The non-monotonic dependence on the interaction should be considered separately in two regimes: 
the gapless regime ($0\le\Delta\le 1$) and the gapful ($1<\Delta$) regime.
There are critical differences between the two regimes in spectral property as well as the ground state.

In the gapless regime, while average ratio $\langle r\rangle$ decreases with increasing disorder $h$ (in the presence of interaction), 
it increases with interaction $\Delta$ (for small disorder).
At $\Delta=0$, $\langle r\rangle\approx\langle r\rangle_{\rm p}$ for an arbitrary $h$. 
In other words, the Poisson statistics appears in the absence of interaction.
This corresponds to AL. 
In the gapless regime, the disorder and interaction dependencies of level statistics are consistent with the results of our previous study~\cite{kudo2004}.

\begin{figure}
\includegraphics[width=8cm,clip]{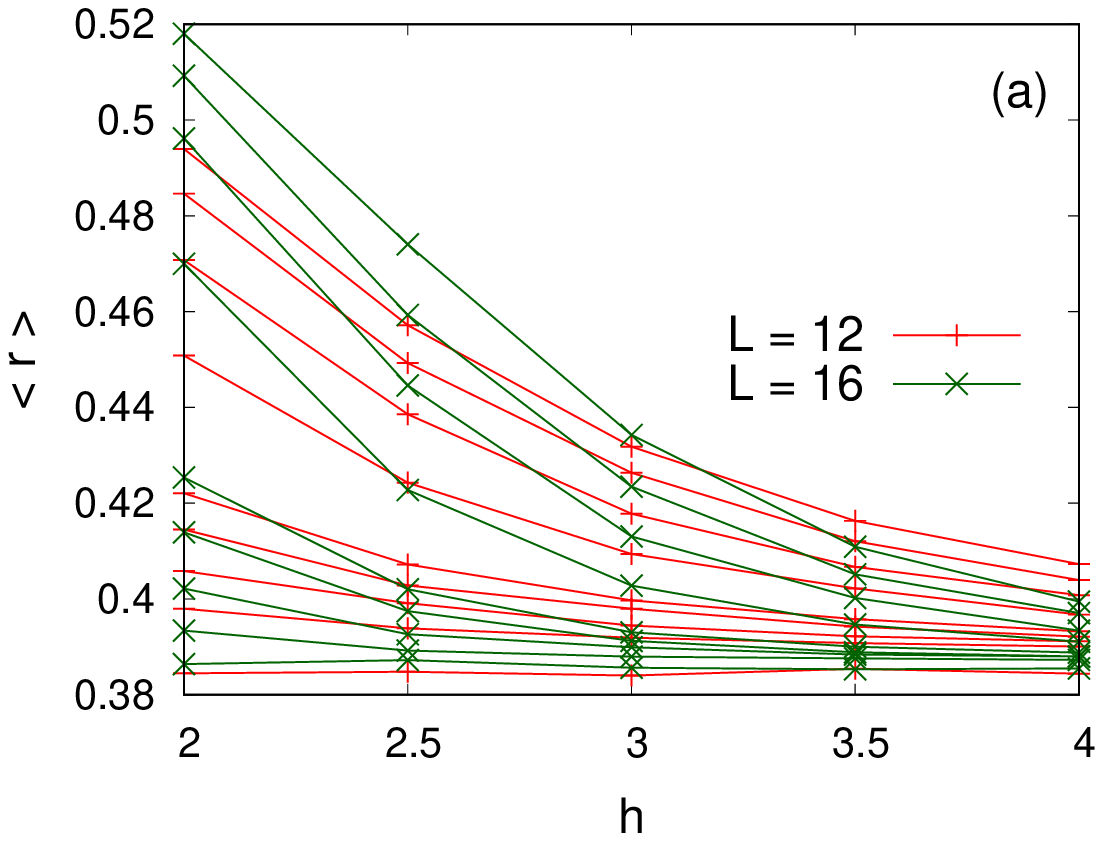}
\includegraphics[width=8cm,clip]{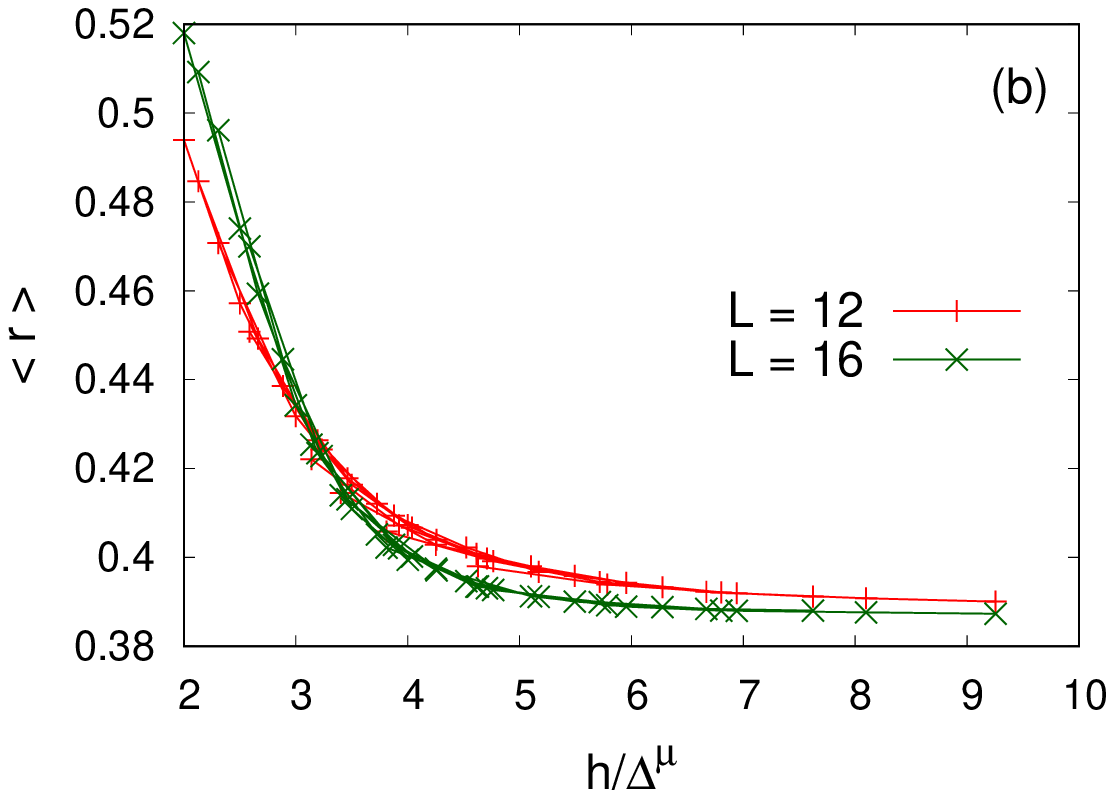}%
\caption{\label{fig:delta_scale} (Color online)  
(a) Average ratio $\langle r \rangle$ with different values of interaction $\Delta$ as a function of disorder $h$ for $L=12$ and $16$.
Lines are for
$\Delta=0,\, 0.05,\, 0.1,\, 0.15,\, 0.2,\, 0.4,\, 0.6,\, 0.8,\, 1$ in order from the bottom, for each $L$.
(b) Rescaled plot of average ratio $\langle r\rangle$ as a function of $h/\Delta^\mu$ with $\mu=0.28$ ($\Delta\neq 0$).
Data of the same system size $L$ collapse into one curve.
The two curves for different sizes cross around $h=h_c$, where $h_c\approx 3.06$ (see text).}
\end{figure}

Next consider the MBL transition in the gapless regime. Here,
the MBL transition is captured as a function of $h/\Delta^\mu$ with $\mu\approx 0.28$, as shown in Fig.~\ref{fig:delta_scale}.
In Fig.~\ref{fig:delta_scale}(a), average ratio $\langle r \rangle$ for different strengths of interaction ($0\le\Delta\le 1$) is plotted as a function of disorder $h$.
The same data, except for $\Delta =0$, are plotted as a function of $h/\Delta^\mu$ with $\mu=0.28$ in Fig.~\ref{fig:delta_scale}(b).
The data of the same system size collapse into a single curve in the rescaled plot.
The point where the two collapsed curves for $L=12$ and $16$ cross is actually consistent with critical disorder $h_c$.
The process leading to these results is described in detail in the next section.

In the gapful regime, however, the MBL transition cannot be captured clearly as a function of $h/\Delta^\mu$, where $\mu$ is expected to be negative.
The nature of the MBL transition in this regime is different from that in the gapless regime.
Actually, in the gapful regime, energy spectra have a lot of large level spacings, namely, energy gaps.
When $\Delta$ is large, because of large energy gaps, the fluctuation of energy levels is relatively small.
Moreover, in the $\Delta\to\infty$ limit, the system corresponds to the Ising model, which is integrable.
In integrable systems, level statistics obeys the Poisson distribution.
This means that energy levels have no correlation.
On the other hand, disorder-induced localization such as MBL and AL also breaks correlation of energy levels and leads to the Poisson statistics.
In the gapful regime, energy-level correlation is diminished by large energy gaps related to integrability rather than MBL.

\section{Finite-size scaling analysis}
\label{sec:scaling}

We estimate exponent $\mu$ in the gapless regime with the procedure of finite-size scaling analysis as follows.
We assume that the data of average ratio $\langle r\rangle$ collapse into a single curve of a finite-size scaling function $g[(h/\Delta^\mu-h_c)L^{1/\nu}]$, where $h_c$ and $\nu$ are the critical disorder and critical exponent, respectively.
In the case of $\Delta=1$, the following procedure corresponds to the finite-size scaling procedure introduced in Ref.~\onlinecite{bertrand2016}.
First, a curve $g_L(h)$ is calculated from the average ratio data with $\Delta=1$, for each $L$.
Next, we calculate the variance over different values of $L$, which is defined by
\begin{equation}
 f_{\rm var}(h)={\rm Var}_L\{ g_L[(h-h_c)L^{1/\nu}]\},
\label{eq:f_var}
\end{equation}
and the average of squared errors over $L$ and $\Delta$, which is defined by
\begin{equation}
 f_{\rm mse}(h)={\rm Avg}_L {\rm Avg}_\Delta
\left\{ g_L[(h/\Delta^\mu-h_c)L^{1/\nu}] 
- \langle r \rangle\right\}^2,
\label{eq:f_dev}
\end{equation}
where $0<\Delta\le 1$.
Note that average ratio $\langle r\rangle$
here is a function of disorder, interaction, and the system size: 
$\langle r \rangle=\langle r(h,\Delta,L) \rangle$.
The curves $g_L[(h/\Delta^\mu-h_c)L^{1/\nu}]$ should coincide with $g[(h/\Delta^\mu-h_c)L^{1/\nu}]$ for appropriate values of $h_c$, $\nu$, and $\mu$ that minimizes the cost function, 
\begin{equation}
 S(h_c,\nu,\mu)=\frac{1}{\Delta h}
\int_{h_{\rm min}}^{h_{\rm max}}[f_{\rm var}(h)+f_{\rm mse}(h)]dh,
\label{eq:cost}
\end{equation}
where $\Delta h=h_{\rm max}-h_{\rm min}$.
We took $h_{\rm min}=2$ and $h_{\rm max}=4$ in the simulations.
Minimization was performed by evaluating the cost function for every combination of $2.5 \le h_c\le 3.5$, $0.5 \le\nu\le 1.0$, and $0.2\le\mu\le 0.4$ at each interval of $0.01$.

\begin{figure}
\includegraphics[width=8cm,clip]{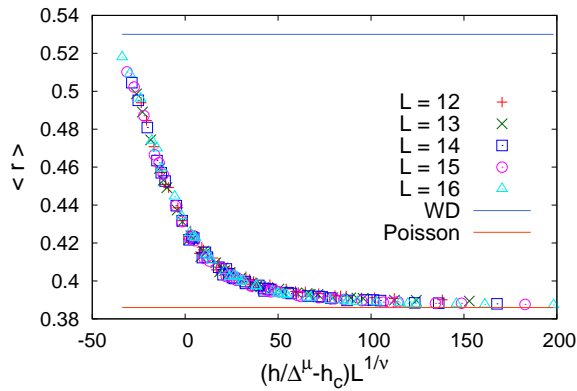}%
\caption{\label{fig:finite_size} (Color online) 
Finite-size scaling of average ratio $\langle r \rangle$, which is also rescaled in terms of interaction ($0<\Delta\le 1$).
Average ratio $\langle r \rangle$ is plotted as a function of $(h/\Delta^\mu-h_c)L^{1/\nu}$ with $h_c=3.06$, $\nu=0.80$, and $\mu=0.28$.
The data with different size and different interaction collapse into a single curve.
The values of $\langle r \rangle$ for the Poisson and Wigner-Dyson statistics, which are indicated as lines, are $\langle r\rangle_{\rm p} \approx 0.386$ and $\langle r\rangle_{\rm w} \approx 0.530$, respectively.
}
\end{figure}

It is clear in Fig.~\ref{fig:finite_size} that a novel data collapse occurs for average ratio $\langle r\rangle$ in the finite-size scaling analysis.
The plots  of the data are rescaled by not only system size $L$ and disorder $h$ but also interaction $\Delta$.
The critical disorder and the critical exponent are estimated as $h_c\approx 3.06$ and $\nu\approx 0.80$, respectively. 
Although the values are slightly smaller than those of Ref.~\onlinecite{bertrand2016},
$h_c\approx 3.35$ and $\nu\approx 0.86$,
they are in a reasonable range of parameter values estimated in other works~\cite{luitz2015,pal2010,serbyn2016}.
The critical exponent of interaction $\Delta$ is estimated as $\mu\approx 0.28$. 

\section{Discussion and conclusions}
\label{sec:discussion}

We suggest that the finite-size scaling in terms of both the interaction and disorder strengths should be interesting from the viewpoint of application of renormalization group arguments to MBL transitions. 
We remark that renormalization group arguments have been applied to study the asymptotic behavior of entanglement entropy \cite{pekker2017}.  
We also remark that the extra exponent $\mu$ is estimated less than 1 in the gapless regime. 
As a result, the effect of change in interaction coupling is weaker than that of disorder.

The rescaled plot in the form of finite-size scaling function in Fig.~\ref{fig:finite_size} clearly illustrates the interaction dependence of the MBL transition.  
It successfully avoids some problems which can occur for a small or large disorder.  
When we study the MBL transition in terms of interaction strength with fixed disorder, 
a naive problem arises: what value of disorder strength we should take.
Obviously, no transition occurs for a large disorder, as shown in Fig.~\ref{fig:map}.
For a small disorder, the change in average ratio $\langle r\rangle$ is seen clearly and easily. 
However, for a very small disorder, spectral properties are affected by the integrability of the system~\cite{kudo2004}.
Here, we recall that the XXZ Hamiltonian with a random magnetic field in Eq.~(\ref{eq:H}) returns to being integrable if the random field vanishes completely.  
We therefore suggest that it should be nontrivial to investigate directly how the MBL transition behaves in the case of a very small disorder.   
Here we remark that 
the MBL transition is usually investigated in terms of disorder strength under a fixed interaction, 
and also that critical properties around the critical disorder are often described well by the finite-size scaling analysis with respect to disorder strength when interaction is fixed.

We have demonstrated how the interplay between interaction and disorder affects level statistics.
The data collapse in Fig.~\ref{fig:finite_size} suggests a nontrivial relation between disorder $h$ and interaction $\Delta$ regarding the MBL transition. 
The MBL transition is observed as a function of $h/\Delta^\mu$ with some exponent $\mu$ in the gapless regime.
Although the nontrivial dependence is evident in numerical results, it has no theoretical support yet.
Theoretical analysis of this dependence will provide deep insights into many-body interaction in quantum systems as well as the MBL transition.

\begin{acknowledgments}
This work is partially supported by JSPS KAKENHI Grant Numbers JP15K05204, JP18K03450, JP18K11333, and the research grant from the Inamori Foundation.
\end{acknowledgments}



\end{document}